# An *In Situ* Light Illumination System for an Aberration-Corrected Environmental Transmission Electron Microscope

Qianlang Liu, Barnaby D. A. Levin, Diane M. Haiber, Joshua L. Vincent, and Peter A. Crozier*

[1]*School for Engineering of Matter, Transport, and Energy, Arizona State University, Tempe, Arizona 85281*

*Corresponding author email: crozier@asu.edu

**Abstract**

In this work, an optical fiber based *in situ* illumination system integrated into an aberration-corrected environmental transmission electron microscope (ETEM) is designed, built, characterized and applied. With this illumination system, the dynamic responses of photoactive materials to photons can be directly observed at the atomic level, and other stimuli including heating and various gases can also be applied simultaneously. The optical fiber enters the ETEM through the objective aperture port, with a carefully designed curvature and a 30° cut at the tip to orient the emitted light upwards onto the TEM specimen. The intensity distributions striking the sample from the broadband and laser sources are both measured, and due to the non-uniform distributions, a procedure has been developed to align the bright spot with the electron optical axis of the TEM. Strong *in situ* laser illumination is demonstrated through the observation of Langmuir evaporation in GaAs, illustrating the phenomenon of optical heating on strongly absorbing material. Photochemical changes in gas are demonstrated through the observation of surface structural changes in $TiO_2$ photocatalytic nanoparticles during exposure to water vapor and UV irradiation.

# 1. Introduction

The importance and advantages of performing *in situ* characterization with transmission electron microscopy (TEM) have been well recognized in recent years, especially in fundamental studies to elucidate functional mechanisms and synthesis pathways (Taheri et al., 2016). By modifying the standard operation conditions of a TEM, direct observations of materials' dynamic responses to selected and controlled stimuli can be achieved. So far, the development of *in situ* techniques has enabled exposure of TEM specimens to various stimuli including gas, liquid, heat, electrical biasing, mechanical stress, and photon irradiation, with the aim of learning materials properties under near-realistic conditions. In particular, one method to introduce gas and even liquid exposure uses a differentially pumped environmental transmission electron microscope (ETEM), where a microreactor is constructed between the objective lens polepieces to maintain higher pressure than other parts of the column (Doole et al., 1991; Boyes & Gai, 1997; Robertson & Teter, 1998; Dai et al., 2005; Levin et al., 2019). Such a differential pumping design imposes fewer constraints on the TEM sample preparation method, and it is potentially more favored for imaging surface atomic structures as well as spectroscopic analysis compared to a windowed cell approach, due to exclusion of electron scattering by the capping membranes (Creemer et al., 2008). Moreover, the latest generation of ETEM incorporates $C_s$ aberration correctors and monochromators to further improve the spatial and energy resolution, which are essential for *in situ* observation of the atomic and electronic structures of nanomaterials and their surfaces (Hansen et al., 2010; Hansen & Wagner, 2012; Boyes & Gai, 2014).

In the general field of photoactivity, functional materials interact with photons for light emission or detection, as well as converting light into electrical current or to accelerate chemical reactions (Diamanti, 2018). For example, in photocatalytic water splitting nanoparticulate light-absorbing semiconductors, with redox co-catalytic surface sites, are employed to convert and store photon energy into the chemical energy of hydrogen molecules (Matsuoka et al., 2007). The atomic and electronic structures of the semiconductor and associated co-catalysts dictate reaction efficiency. However, it is known that the original catalyst structures may alter or transform under reaction conditions. In fact, Zhang et al. have observed that, both *in situ* and *ex situ*, the initially crystalline surface structure of anatase nanoparticles converts to a disordered phase one to two monolayers thick, accompanied with formation of $Ti^{3+}$ cations upon exposure to light and water (Zhang et al., 2013). In addition, structural evolutions of co-catalysts including loss of metallic Ni from Ni-NiO core-shells, or coarsening of initially highly dispersed Pt nanoparticles are found to be related to reactivity drop (Liu et al., 2015; Zhang et al., 2015, 2019). These findings illustrate the importance of *in situ* characterization of photocatalysts under near-reaction conditions is of great value in revealing the reaction, aging and deactivation mechanisms. Integrating an *in situ* illumination system onto an aberration-corrected ETEM is therefore highly desired, in order to understand the responses of photoactive materials to photon and other stimuli at the atomic or nanometer level.

Several approaches that allow exposure of TEM samples to photons have been investigated over the years. For example, a light illumination system built by Yoshida et al. was coupled to a high-resolution TEM to study the decomposition of hydrocarbons on catalytic $TiO_2$ at the atomic level (Yoshida et al., 2004). Cavalca et al. developed two types of TEM specimen holders where a lens-based and a fiber optics-based illumination system was integrated on the holder respectively (Cavalca et al., 2012). Recently, a novel optical TEM holder with a multimode fiber, a piezo-motor driven metal tip, and a bias voltage of around 10 V was developed by Zhang et al. for measuring the optoelectronic properties of nanostructures (Zhang et al., 2014). A parabolic mirror based system is designed and used to focus external light onto TEM specimen as well as collecting the sample emitted photons for Raman or cathodoluminescence (CL) signals (Picher et al., 2015). Other approaches have been developed to employ laser sources and fiber optical coupling and lenses to enable intense photon irradiation to perform *in situ* heating (Rossouw et al., 2013; Wu et al., 2018). Ultrafast electron microscopy often employs pump-probe approaches in which short laser pulses are employed to trigger phase transition which are subsequently followed in time with short electron pulses (LaGrange et al., 2012).

Here we have designed a simple fiber optic-based device which is independent of specimen holder, which can be integrated into the objective aperture port of an aberration-corrected ETEM (FEI Titan 80-300), directing photons from either a wide band light source or a more intense laser source onto the specimen. This system allows simultaneous exposure of the sample to high intensity photon irradiation, various gas species with pressures up to 20 Torr, as well as heating, cooling, biasing or mechanical stimulus since specially designed *in situ* specimen holders can still be employed. The design concept is similar to one we previously developed for *in situ* illumination system on an uncorrected ETEM (FEI Tecnai F20) (Miller & Crozier, 2013). Compared to the Tecnai system, having the imaging $C_s$ corrector as well as a monochromator makes it possible to reveal further detailed information on the structural changes taking place under light illumination and reactive gas environments on nanoparticles and surfaces.

## 2. Methodology: Design Criteria and Considerations

Two primary design criteria are set for this *in situ* illumination system: (A) the desired optical functionality needs to be realized, *i.e.*, sufficiently high photon intensity needs to be achieved at the sample area, and (B) integrating the illumination system into the ETEM should not significantly affect the microscope capabilities, which includes introducing reactive gases into the environmental cell in addition to obtaining high-resolution images and good quality electron energy-loss spectroscopy (EELS) data. In short, criterion (A) deals with the functionality of the system, while criterion (B) deals with compatibility. To achieve these two criteria, various factors and issues need to be considered, and are discussed as follows.

*Considerations related to functionality: Criterion (A)*

The first step involved in realizing the functionality of the *in situ* light illumination system is to decide how to direct and focus light from a light source onto the TEM sample. A fiber optic based approach has been used previously in a Tecnai ETEM system (Miller & Crozier, 2013). Compared to a lens-based design, a fiber optic approach offers greater simplicity and flexibility, and so it is chosen here for the current system. To allow for versatility and the choice of any *in situ* specimen holder with heating, cooling, biasing (or other capabilities), the fiber optic design is independent of the stage and delivered through a separate port on the column close to the sample; here, we selected the objective aperture port. Directing the light from the source, through the port, and onto the sample is a challenging task, largely because of geometric constraints set by (1) the spatial limitations of the objective aperture port, and (2) the limited space between the pole pieces and the specimen stage (see, *e.g.*, **Figures 1 and 2**). The objective aperture port is comprised essentially of two bores with diameters of 10 and 6 mm. The space between the upper and lower objective pole pieces in the Titan microscope is roughly 5.4 mm. For *in situ* experiments which employ a furnace-based heating holder, the TEM specimen resides within an opaque cylindrical furnace that has a depth of ~2.4 mm and an inner diameter of ~3.4 mm. All of these geometric constraints, along with the maximum allowable curvature of the fiber, limit the maximum angle at which the fiber can approach the sample.

The spatial distribution of the photon intensity reaching the sample area is another factor that needs careful consideration in order to achieve the desired functionality. Ideally, it is desired to have high intensity light uniformly illuminating the entire sample area, which can be as large as a circular grid with a diameter of ~3 mm. In practice, however, the light emitted from the fiber is not collimated, and has a non-uniform intensity distribution (Miller et al., 2012). In other words, a bright spot containing relatively higher photon intensity is expected to be present near the center of the spatial distribution of the light intensity. To ensure that sufficient photon intensity can reach the sample area, the light source needs to have high brightness or radiance. Furthermore, it is essential to be able to adjust the position of the fiber once it is installed on the microscope, so that the bright spot can be aligned with respect to the electron optical axis of the TEM. It should also be noted that the optical fibers and fiber connectors used to direct light should be able to withstand the possible damage caused by high energy photons.

*Considerations related to compatibility: Criterion (B)*

Integrating the fiber optic based light illumination system onto an ETEM involves several steps. First, seals need to be made so that after installation of the system, the microscope chamber can be pumped down to the desired vacuum level (~$10^{-7}$ Torr in the objective pole pieces area). This also means that the components sitting inside the microscope vacuum should be vacuum compatible and that residual air in these components should be easily pumped out. In addition, it is desired that the components can be baked at 100 – 150 ºC before installation to suppress contamination to the microscope chamber. Second, the components need to be stable when exposed to various reactive gases, such as water vapor, $O_2$, $H_2$, etc., during *in situ* experiments. Therefore, they should not be made of materials that are easily hydroxylated, oxidized, or reduced. Third, the materials should not degrade the imaging and spectroscopy performance, which means they should not interfere with the strong electromagnetic field (usually ~1-2 T) in the pole piece area, and should be able to conduct away the spurious, scattered, and secondary electrons to eliminate any charging effect.

*Other considerations*

In addition to the considerations related to the system's functionality and compatibility, it is also important to ensure that no significant leakage of X-rays generated from electron-solid interactions is present after installing the system through the objective aperture port. Moreover, the system should be designed in a way that it can actually be fabricated at an affordable cost. Also, the assembling/disassembling of various components, as well as installation/dismount of the system should not be too difficult.

## 3. Results and Discussion

*Design of fiber optic light illumination system*

Each of the considerations discussed above has been addressed in the current system which was designed to be installed in an FEI Titan ETEM. AutoCAD was used to construct an original design of the fiber optic holder, which was then fabricated by the mechanical instrument shop at Arizona State University. **Figure 1a** shows a design drawing of the assembled device, located at the optimum position with respect to a tilted sample in between the objective pole pieces. The geometric constraints set by the bores of objective aperture port are indicated with dashed red lines. A picture of the actual fabricated device is also shown in **Figure 1a**, below the design drawing. The device has five major components: a fiber holder, a specially designed optical fiber (fiber 2), a fiber adapter, a vacuum feedthrough, and a screw cap, as illustrated in the exploded view drawing in **Figure 1b**. Detailed design drawings of the fiber holder and the fiber adapter are provided in **Supplemental Figure S1**.

The main function of the fiber holder is to support the optical fiber while maintaining a certain contour/bend so that the tip of fiber points upwards leading to the TEM sample. A maximum obtainable upward angle of ~9° is realized by carefully designing a contour/bend in the holder, as seen in **Figure 1b**. It should be mentioned that the contoured portion of the holder sits inside the microscope high vacuum while the axial symmetric portion does not; an O-ring is placed at their joint to form a vacuum seal (vacuum seal A in **Figure 1a**). A 3-mm diameter hole is drilled along the center axis of the symmetric portion of the holder for inserting the optical fiber, which itself has an outer diameter of 0.8 mm. The bend in the contoured portion has a radius of 25 cm, which is the largest allowable curvature to prevent inefficiencies in light transmittance. A copper tubing is inserted to follow the contour, and the fiber goes through this tubing to achieve the desired bend. Several small apertures are drilled in the copper tubing and the body of the contoured portion of the holder to rapidly pump out trapped residual air.

The optical fiber in the holder (fiber 2) is specially designed to meet important design criteria outlined in Section 2. The fiber has a refractive index of 1.47 and is suitable for transmitting UV light (*i.e.*, solarization-resistant). In contrast to the fiber located outside of the microscope (fiber 1, see **Figure 1a**), which has a standard polymer-based protective buffer and jacket, this fiber is coated with an aluminum buffer which minimizes outgassing and avoids degradation from reactive gasses. The aluminum buffer is also conductive and non-magnetic, both of which are desirable in the electron microscope. The fiber tip is cut at a 30° angle. As shown in **Figure 2**, the angled surface at the end of the fiber directs the light emitted from the tip, by refraction, upwards, and onto the TEM specimen. When the fiber is inserted at an optimum position and the TEM sample is tilted 30° towards the fiber, the light cone (shown in green in **Figure 2**) emitted from the tip strikes the TEM sample without being blocked by the furnace of a heating holder. The other end of this fiber is terminated with a ferrule that is a standard SMA 905 fiber connector (shown in black in **Figure 1b**). The ferrule is made of stainless steel so that the entire fiber can be baked in the $100 - 150$ °C range.

The 3 mm hole drilled in symmetric portion of the holder is too small to accommodate the ferrule from the fiber, thus a fiber adapter has been made which has a 9 mm hole drilled along its center axis. A rotatable 1.33 CF flange is placed at the end of the adapter to interface with the vacuum feedthrough. The interface is noted as vacuum seal B in **Figure 1a**. Having a rotatable flange is essential, as the 30° cut at the tip of the fiber needs to be oriented so the light emitted from it is refracted upwards and onto the TEM specimen. In addition to forming a second vacuum seal, another purpose of the vacuum feedthrough is to couple fiber 2 with fiber 1 using the two SMA fiber connectors on both sides of the seal. The vacuum feedthrough is selected to be extreme solarization-resistant (XSR) to suppress inefficiency in transmitting the UV light.

Besides the fiber adapter and vacuum feedthrough, a hollow screw cap is also necessary to fix the assembled device onto the objective aperture port, and to compress the O-ring at vacuum seal A. An X-ray leakage test has been performed after installing this device onto the microscope, and the detected leakage is below the maximum allowable limit, confirming that the emitted X-rays can be sufficiently absorbed by the fiber holder. A summary of the materials and a description of the items that comprise the components of the fiber holder is presented in **Table 1**.

*Light distribution and fiber alignment*

Either a broadband light source or a laser source can be utilized in the current optic fiber-based system. The flexibility in light source permits different irradiation conditions to be applied in the study of various photoactive materials and reactions of interest. The broadband light source is an Energetiq EQ-99 laser-powered source providing high radiance over the wavelength range of ~200 to 800 nm, and with the current design, yields an incident power density between $10^2$ to $10^3$ mW/cm$^2$ at the sample. A detailed description of this source and associated operating principles can be found in (Miller et al., 2012). In addition to the broadband light source, a more intense laser source is also available, with the benefit of expediting certain photoreactions. The laser source is a 405 nm (3.06 eV) fiber-coupled laser purchased from CivilLaser (CivilLaser, 2020). The FWHM around the center wavelength is 12 nm. In this system, the laser power is controlled with an adjustable current supply that can achieve a maximum total power output of 4 W. The output power density incident on the TEM sample has been characterized for a range of current supply settings, and a calibration plot is provided in **Supplemental Figure S2**. In general, the laser power density at the TEM specimen can be varied up to a maximum value of $10^5$ mW/cm$^2$.

The spatial distribution of the light intensity incident on the TEM specimen is important to understand and characterize. The distribution can be measured outside of the TEM with an optical microscope, using a method described in references (Miller et al., 2012; Liu, 2017) and a typical normalized light intensity distribution is shown in **Figure 3**. The contours correspond to 10% intervals within the normalized intensity profile, which shows that for both sources the light intensity drops quickly away from the brightest spot. The bright spot of the profile may be defined to be the area where the intensity remains within 90% of the maximum (*i.e.,* the dark red region in **Figures 3a** and **3b**). The bright spot covers $0.44 \times 0.58$ mm$^2$, or about 3.6 % of the area available on a standard 3 mm TEM grid. The bright spot dimensions and a circle representing a standard 3 mm TEM grid are shown in **Figure 3** for reference.

Given the small coverage area of the bright spot, it is essential to precisely align the fiber tip to ensure that the center of the bright spot is concentric with the optic axis of the TEM. The fiber may be controlled with the bellows and rotors originally designed for the objective aperture rod. With respect to the axis in **Figure 2**, the optic fiber can be inserted or retracted (x-direction) as well as laterally shifted (y-direction) with a precision of 0.1 µm. Vertical movement (z-direction) is not available but also not required to achieve the desired optical functionality. Although precise control of the fiber position is possible, it is still challenging to align the fiber with respect to the specimen because the location of the fiber tip cannot be directly observed once the fiber is installed into the TEM. An alignment procedure has thus been developed whereby the optic fiber is used to detect light emitted by a sample of phosphorescent particles loaded onto the TEM grid. A schematic of the alignment procedure is shown in **Figure S3.** In brief, green phosphor nanoparticles (ZnS based P22) are loaded onto a TEM grid which is then placed at the eucentric height and tilted 30º towards the fiber. Upon irradiation by the electron beam, a phosphor particle will emit photons isotropically. A portion of the emitted light may enter the optic fiber and be detected by a photosensor system that has been installed at the end of the fiber in place of a light source outside the microscope. The main component of the photosensor system is a photomultiplier tube (PMT) made by Hamamatsu Corporation, chosen here to be most sensitive to green light (H6780-02 module). The output voltage from the PMT increases with increasing incoming photon intensity or number, and so the position of fiber 2 can be aligned by moving it until maximum voltage is achieved. Finally, *in situ* light illumination produces a slight shift in the entire TEM image (see, *e.g.*, Supplemental **Figure S4**). In this case, the image shift may have changed the electron path length through the sample, leading to the small difference in signal intensity.

*Performance of the light illumination system and impact on imaging*

The performance of the installed light illumination system was investigated, and the system's impact on imaging during light illumination was examined. Various photoactive specimens of interest were employed to investigate the system's performance. Both the 405 nm laser and the broadband light source were used during testing as discussed below.

The impact of light illumination on the image resolution was investigated by imaging an ion milled GaAs thin film specimen under conditions of increasing light intensity with the 405 nm laser source. The aberration-corrected Titan ETEM was used to acquire images in a negative spherical aberration coefficient imaging mode at 300 kV. The AC-TEM image of the GaAs thin film acquired when the laser is off is shown in **Figure 4a1**. A diffractogram of this image is displayed below it in **Figure 4b1**, which shows information transfer out to 0.74 Å, as indicated by the dashed yellow circle. As the laser light illumination system is turned on and the sample begins to be irradiated with 405 nm light of increasing intensity (**Figures 4a2 – 4a4**), atomic-columns remain clearly resolved in the image. Here the laser power density striking the sample in each image is kept to < 1 W cm$^{-2}$ in order to minimize laser beam-driven specimen damage; the power for each light illumination condition is provided in the inset of every respective image. The associated diffractograms (**Figures 4b2 – 4b4**) show a slight attenuation of the spots in the diffractogram at the highest spatial frequencies, but overall information close to 1 Å can still be seen when the laser is on and illuminating the specimen at close to 1 W/cm$^2$. This is an important result since it demonstrates that the image resolution of the AC-TEM is not significantly compromised during *in situ* light illumination.

*Application of high intensity in situ illumination to GaAs*

The *in situ* illumination system was applied to GaAs specimen to demonstrate the high power capability available with the 405 nm laser light source. GaAs is a direct bandgap semiconductor with a bandgap of 1 eV and relatively high absorption coefficient at 3 eV (the laser energy). **Figure 5a** shows a TEM image of the GaAs thin film after brief (3 – 5 second) exposure to the 405 nm laser at a power density of 19 W/cm$^2$. Careful inspection of the image reveals the appearance of dark patches of contrast 10's of nm in size most visible along the thickness fringes which run parallel to the surface of the thin film. **Figure 5b** shows a TEM image of the same region of the sample after brief laser exposure at a power density of 33 W/cm$^2$. Severe chemical and structural transformations have occurred after increasing the laser power density to 33 W/cm$^2$, as evident by the surface pitting and evolution of numerous dark and round patches of contrast. Such severe transformations were not observed under condensed and prolonged electron beam illumination, moreover, such behavior was also observed at regions in the sample that had never been illuminated with the electron beam but had been exposed to the laser. Thus electron beam damage effects are not the cause the observed structural transformations. Instead, these observed transformations can be attributed to the interaction between the GaAs and the incident 405 nm laser light, an effect known as optical heating, which results from thermalization effects from photo-excited electrons (Meyer et al., 1980).

Energy dispersive X-ray (EDX) spectroscopy was used to investigate the nature of the chemical and structural transformations that took place in the GaAs thin film during high power laser light illumination. **Figure 6a** shows EDX spectra taken of the GaAs thin film specimen before (blue line) and after (red line) exposure to the 405 nm laser at a power density of 33 W/cm$^2$. A comparison of the two spectra shows that the ratio of Ga:As in the sample has increased after high power laser light illumination. **Figure 6b** shows an EDX spectrum taken from one of the dark patches of contrast visible in the TEM images presented in **Figure 6b**. The spectrum demonstrates that the dark globular objects show a large enrichment of Ga, as evidenced by the prominent peaks corresponding to Ga relative to those corresponding to As. Taken together, these data suggest that the energy deposited into the GaAs specimen from absorbing the 405 nm laser light has caused the As to sublime and the Ga to melt into globular clusters. These observations are consistent with Langmuir evaporation of GaAs, which has previously been observed to occur at ~630 °C in a photoemission electron microscope (PEEM) (Tersoff et al., 2010). Thus the high power density of the laser has resulted in energy being absorbed into the GaAS giving rise to optical heating to temperatures about 630°C. This shows that even this simple optical fiber system with no focusing can generate significant sample heating at least in materials like GaAs where the absorption characteristics are well matched to the photon wavelength.

*Anatase Phase Changes in the Presence of Light and Water Vapor*

To investigate the system performance for exploring photo-reactions, we investigated the reaction between anatase nanocrystals, water vapor and light. Anatase ($TiO_2$) is an important photocatalytic material which is inexpensive and generally stable under photo-reactions conditions (Fujishima et al., 2008). However, our previous work on an FEI Tecnai microscope showed that amorphous surface layers form under photoreaction conditions (Zhang et al., 2013). We initially set out to reproduce those results on the Titan ETEM.

Pure crystallized and shape-controlled anatase nanoparticles were synthesized using a hydrothermal method described elsewhere (Zhu et al., 2005; Qixin et al., 2010). The anatase powder was directly dispersed over a 3-mm Pt mesh support for the *in situ* experiments because a standard carbon film is not stable due to etching when exposed to light and water, whereas Pt mesh does not degrade under these experimental conditions. A Gatan hot stage was used as the specimen holder and the sample was heated to 150 °C to suppress carbon contamination that may occur in non-UHV electron microscopes during long irradiation times.

To expose the anatase nanoparticles to conditions relevant to vapor phase photocatalytic water splitting, the sample is tilted 30° towards the fiber and the Energetiq light source is employed with the maximum aperture size selected and with no optical filter inserted, i.e., the sample is illuminated with both UV and visible photons (i.e. 200 – 800 nm) and a total incident power on the sample of ~ 1400mW/cm². To introduce water vapor into the environmental cell, a glass bottle containing liquid DI water is connected to a leak valve on one gas inlet of the cell. The liquid water is partly vaporized as the vapor pressure of water at room temperature is approximately 20 Torr. By adjusting the opening of the leak valve, the water vapor pressure in the environmental cell can be controlled. In this experiment, the water vapor was first kept at 1 Torr for 11 h then increased to 8 Torr for 3 h while the sample is exposed to light throughout.

The Titan microscope is operated at 300 kV and anatase surface structures were examined before and after the 14 h *in situ* exposure to light and water. To minimize radiation damage (which occurs rapidly in water vapor), the sample were examined after exposure using low electron dose, high vacuum condition ($\lesssim 10^{-4}$ Torr) and no photon irradiation. It should be noted that both electrons and photons can generate electron-hole pairs in the anatase which may result in the formation of surface active sites for water dissociation (Wang et al., 1994; Zhang et al., 2013). Therefore, a relatively low electron dose rate (~300 $e^-/Å^2 \cdot s$) was used for acquiring the high resolution images to suppress the effect of electron irradiation on the structural changes taking place in anatase. This electron dose rate, or the corresponding incident electron power density, is much larger than the incident photon power density from the Energetiq source (Zhang et al., 2013). The method to calculate the energy transferred to the TiO2 by the electron beam and the incident photon is describe in detail in Zhang et al. For the low electron dose rates, the per unit area energy transferred from the electron beam to the sample after exposure can be estimated from the electron energy-loss spectrum to be ~54 $J/cm^2$ (assuming it takes about 20s to focus the sample before imaging). For the white light source, only incident photons of energy greater than the bandgap (~ 122 mW/cm² incident) are absorbed and the absorption coefficient and average particle size gives an effective absorbed power density of ~ 55 mW/cm² resulting in a total absorbed surface energy from the light is ~2777$J/cm^2$ during a 14 hour exposure. This calculation confirms that most of the energy absorbed by the sample is primarily from photon irradiation. As a further check on electron beam damage effect, anatase particles that had not been previously exposed to electrons before the *in situ* treatment, were inspected to further verify if the changes were due to light irradiation.

**Figure 7a** and **b** display initial and final images of an anatase nanoparticle showing a stepped and vicinal (101) surface before and after exposure to light and water vapor. Initially the particle surfaces appear clean, almost bulk-terminated but after 14 h *in situ* exposure to light and water vapor, an order-to-disorder transformation is observed on the surfaces of this particle and adjacent particles. The order-to-disorder transformation is identical to our previous observation and is associated with hydroxylation of the oxide surface triggered by the presence of UV induced oxygen vacancies on the surface (Zhang et al., 2013). The disordered layer thickness varies across the surface and this heterogeneity may be related to variations in the initial surface structures. For example, surfaces initially containing higher concentrations of steps or undercoordinated sites may result in higher degrees of hydroxylation possibly due to lower oxygen vacancy formation energies and lower activation energies for dissociative adsorption of water molecules at these sites.

During this testing period we were fortunate to have temporary access to a Gatan K2 direct electron detection system. This is particularly powerful for radiation sensitive systems because it allows us to go significantly lower in electron dose while maintaining reasonable signal-to-noise ratio (SNR). This detector system enabled us to perform higher spatial resolution measurements of the changes in anatase after exposure to water and light. **Figure 8** shows a high resolution image of an anatase nanocrystal which is oriented along the (010) zone axis before exposure to water and light. For this work, the aberration corrector was set to provide a negative $C_s$ imaging condition ($C_s$ ~ -20 μm) giving so-called white contrast for atomic columns and enhanced contrast from oxygen. In this projection, the brighter dots correspond to Ti-O columns (arranged in staggered rows of dumbbell pairs) and the fainter dots to oxygen columns. The initial imaging was performed in vacuum in the absence of light with a nominal electron dose rate of 250 $e^-/Å^2$ s. The process of focusing and adjusting the imaging conditions was estimated to be roughly 20 s resulting in a total dose of around 5000 $e^-/Å^2$ s. The nanocrystal becomes thinner towards the surface giving rise to a weaker surface signal.

**Figure 9a** and **b** show the tip of the same nanoparticle of **Figure 8** just before and after exposure to $10^{-2}$ Torr of water and 10 hours of under the same conditions as **Figure 7**. To minimize electron beam damage in the water environment, **Figure 9b** was recorded with an electron dose of 72 e$^-$/Å$^2$ s and the image was acquired within a few seconds. Although the SNR of **Figure 9b** is significantly lower than 9a, it is clear from the image that structural change has taken place during light irradiation in the presence of water vapor. One obvious change is that the average projected spacing of the Ti-O dumbbell has shrunk from 2.38 Å (one quarter of the c-axis spacing of 9.51Å) to about 1.9 Å. Although this is almost a 25% contraction in the dumbbell spacing, the c-axis periodicity is not significantly changed. since the gap between staggered rows of dumbbells has increased from 2.38 to 2.84 Å. Moreover, the dumbbells which are precisely oriented along the c-axis in Figure 9a show random tilts of up to 15º away from the vertical in **Figure 9b** suggesting the introduction of structural heterogeneity. The poor SNR ratio at the surface significantly degrades the visibility of the surface structure but there is little evidence for long-range periodicity, consistent with the presence of a disordered surface layer as observed in **Figure 7b**.

A complete understanding of the changes taking place in the anatase during exposure to water and light is beyond the scope of the current manuscript but here we provide some possible speculations based on the observations. As shown previously, under UV irradiation, oxygen vacancies can form at the surface providing a path for interaction with water leading to hydroxylation and then associated disordered layer formation (observed in **Figure 7b)**. The nanoparticle in **Figure 9** terminates with a (001) surface and is flanked with vicinal surfaces consisting of closely spaced (001) steps. The steps are mostly composed of dumbbell pairs giving an overall step height of about 4.74 Å. The steps appear to provide an intercalation pathway leading to a 0.4 Å expansion in the dumbbell layer spacing. The chemical identification of the intercalating species is not known but the 2.84 Å can easily accommodate hydroxyl species. We hypothesize that the initial hydroxylation takes place at the step edge and gradually intercalates between the rows of dumbbells (illustrated schematically in **Figure 9c**). The replacement of oxygen anions (charge -2) with hydroxyl species (charge -1) may lead to increased repulsion between the Ti cation giving rise to the increase layer spacing observed. The tilting of the dumbbell pairs suggests that the intercalation is not uniform or symmetric. Further work is required to verify this interpretation and to understand how such changes may affect catalytic activity.

## 4. Conclusions

In summary, an optical fiber-based *in situ* light illumination system has been designed and built for an aberration-corrected environmental TEM (FEI Titan). The system can utilize either a broadband light source or a laser source. A specially designed fiber with an angled cut at the tip was inserted into the microscope through the objective aperture port and used to guide light onto a TEM sample tilted 30° towards the fiber. The fiber must be bent in order to direct light upwards onto the specimen. A fiber holder and an adapter were designed and fabricated to support the fragile fiber and to provide the necessary curvature for illumination. The spatial distribution of the light coming from the fiber tip and incident onto the TEM specimen was measured, and it was found that the dimensions of the brightest spot are much smaller than a standard 3 mm TEM grid. Therefore, an alignment procedure was developed using a phosphorescent specimen and a photosensor system, in order to coincide the position of the bright spot with the electron optical axis of the TEM. The power density striking the sample has also been measured; for the broadband source, the power density is between $10^2 - 10^3$ mW/cm$^2$, whereas for the laser source employed here, the power density incident on the specimen can be up to $10^5$ mW/cm$^2$. The high power available with the laser was demonstrated using a light-absorbing GaAs test specimen. Optical heating from the laser induced rapid Langmuir evaporation of the GaAs, which typically occurs at temperatures in excess of 630 °C, indicating a high level of energy transfer to the specimen.

The set-up was employed to simulate aging of photocatalytic anatase nanoparticles in water vapor during exposure to moderate broadband light irradiation for periods of up to 14 hours. By maintaining a sample temperature of 150°C, we avoid the deleterious effects of carbon contamination that often occurs during extended photon irradiation of TEM samples under non-UHV conditions (i.e. the vacuum in most electron microscope columns). In agreement with earlier work, we observed the formation of a disordered layer on the surface corresponding to hydroxylation showing that the system can reproduce photocatalytic conditions. By employing low electron dose with a direct electron detection system, we were able to investigate the changes that take place in anatase at much higher spatial resolution while controlling electron beam damage that can easily destroy hydroxide layers. These new observations show that significant structural distortion occur at subsurface sites in the anatase. The expansion of the rows of Ti dumbbell associated with surface step edges suggest that intercalation of hydroxl species may take place under reaction conditions.

## 5. Acknowledgements

QL, BL, DH and PAC acknowledge support from the US Department of Energy (DE-SC0004954) and JV and PAC acknowledges support from the National Science Foundation (NSF CBET 1604971). The authors also thank Arizona State University's John M. Cowley Center for High Resolution Electron Microscopy for microscope resources and Richard Flubacher from the mechanical instrument shop at ASU for vital assistance on manufacturing the device needed for the *in situ* light illumination system.

**Tables and Figures**

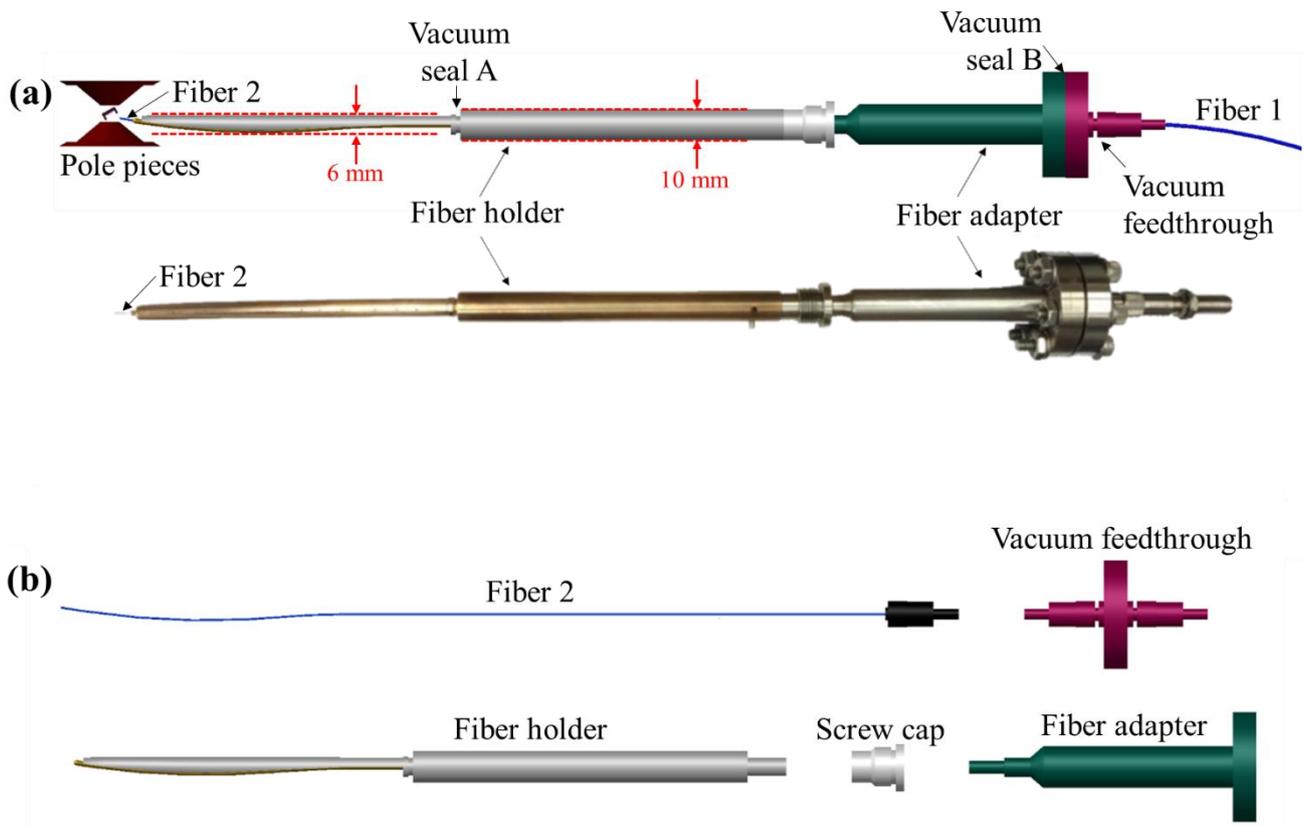

**Figure 1. (a)** A design drawing of the optic fiber-based light illumination system. Light is directed from the light source to the sample area through fiber 1 and then fiber 2. Vacuum seals are made at points A and B so that fiber 2 and certain parts of the system are under the high vacuum of the microscope. Geometric constraints set by the 10 and 6 mm diameter bores of the objective aperture port are indicated with dashed red lines. **(b)** A design drawing in an exploded view to emphasize the major components of the optic fiber-based light illumination system.

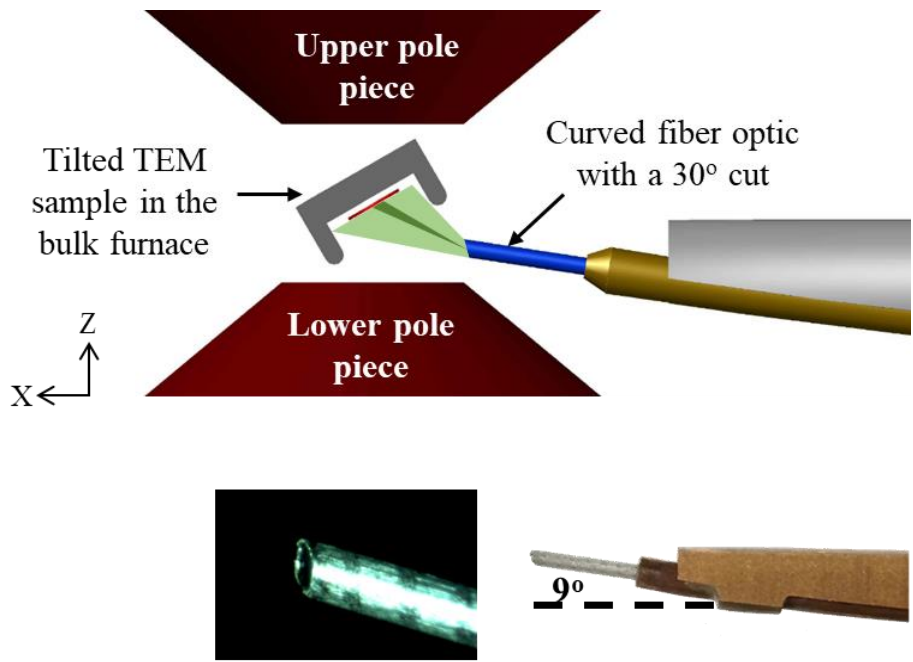

**Figure 2.** A zoom in design drawing focusing on the fiber tip region, showing high intensity light striking the electron optical axis (also the center of the tilted sample grid) when cut tip of the fiber is at an optimum position. A picture of the fiber tip region of the fabricated device and an optical micrograph of the fiber tip showing the aluminum buffer are also displayed.

**Table 1.** A summary of the materials and a description of the items that comprise the device.

| Item | Materials | Description |
|---|---|---|
| Fiber 1 | Silica, polymer, stainless steel ferrule | Ordered from Ocean Optics<br><br>600 μm, solarization-resistant (SR)<br><br>2 m, SMA-905 |
| Fiber 2 | Silica, aluminum, epoxy, stainless steel ferrule | Specially ordered from Ocean Optics<br><br>600 μm, Al buffered, UV/SR<br><br>286.55 mm<br><br>SMA-905 to 30° cut |
| Fiber holder | Phosphor bronze, Ag soldering, Cu tubing | Fabricated at the machine shop at ASU |
| Fiber adapter | Stainless steel | |
| Screw cap | | |
| Rotatable flange | | Fabricated at the machine shop at ASU<br><br>1.33 CF |
| Vacuum feedthrough | Stainless steel, silica | Ordered from Ocean Optics<br><br>600 μm<br><br>1.33 CF flange, XSR |
| O-rings | Viton | One at vacuum seal A<br><br>One between the adapter and the holder |

| Cu gasket | Cu | Creates a seal between the 1.33 CF rotatable flange and the vacuum feedthrough |

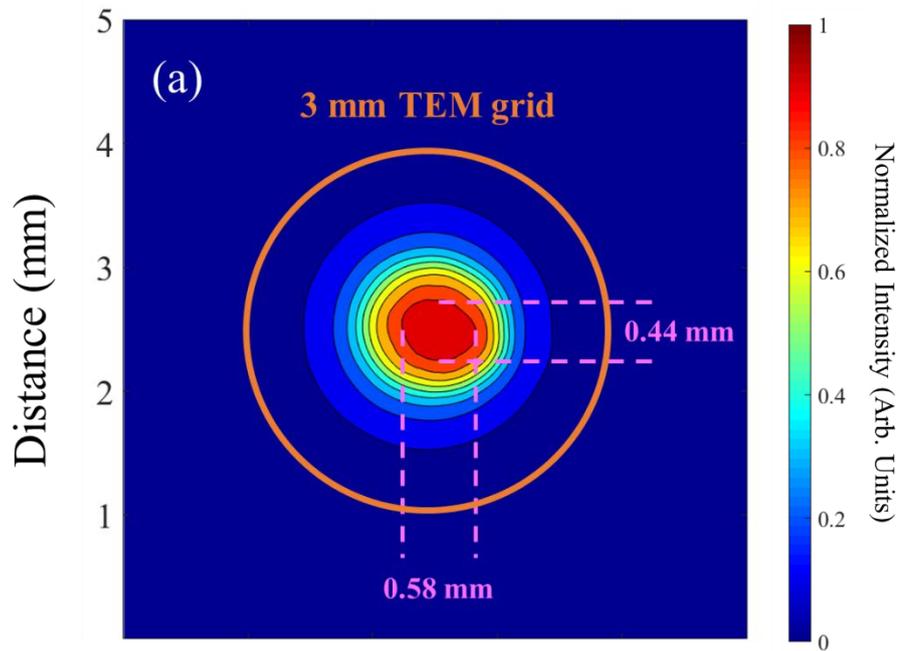

**Figure 3.** The spatial distribution of the normalized light intensity incident on a TEM specimen for the broadband light source. The dimensions of the brightest spot of > 90% maximum intensity are indicated for both sources, and an orange circle representing a standard 3 mm TEM grid is provided for reference.

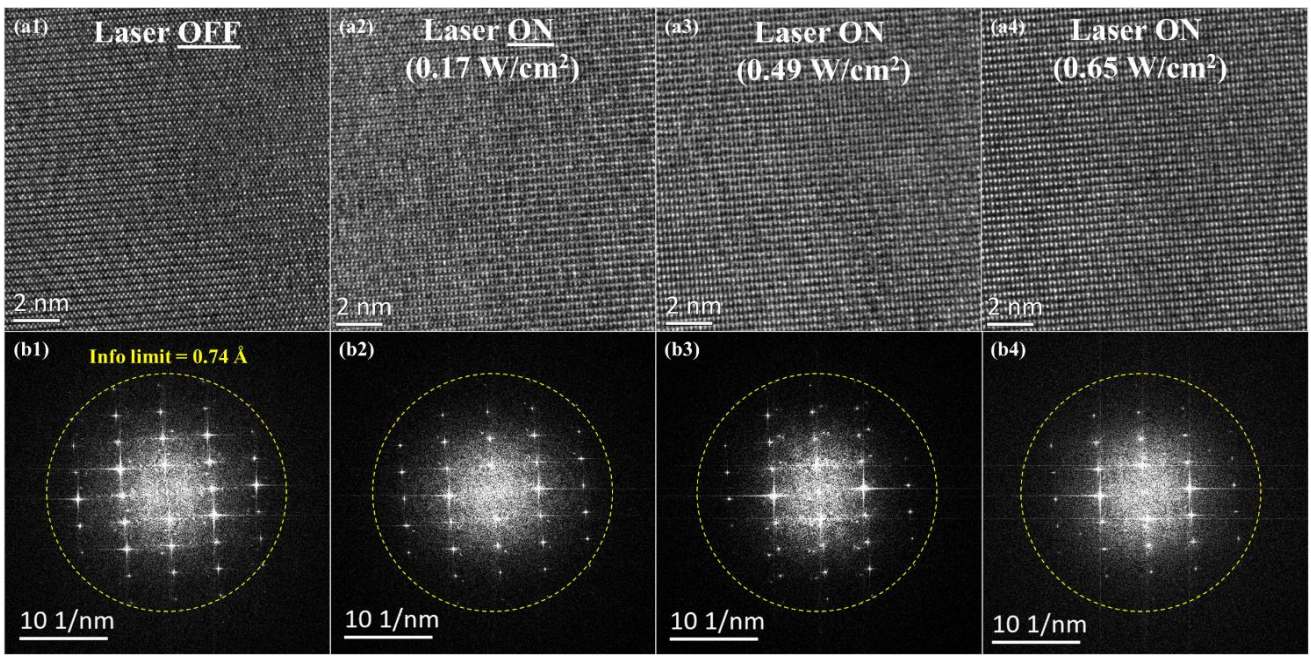

**Figure 4.** High-resolution AC-TEM images of GaAs thin film specimen with **(a1)** the *in situ* laser off, and **(a2 – a4)** the laser on at progressively higher power settings. Figures **(b1 – b4)** show corresponding diffractograms from each image with the instrument's 0.74 Å information limit marked by a dashed circle.

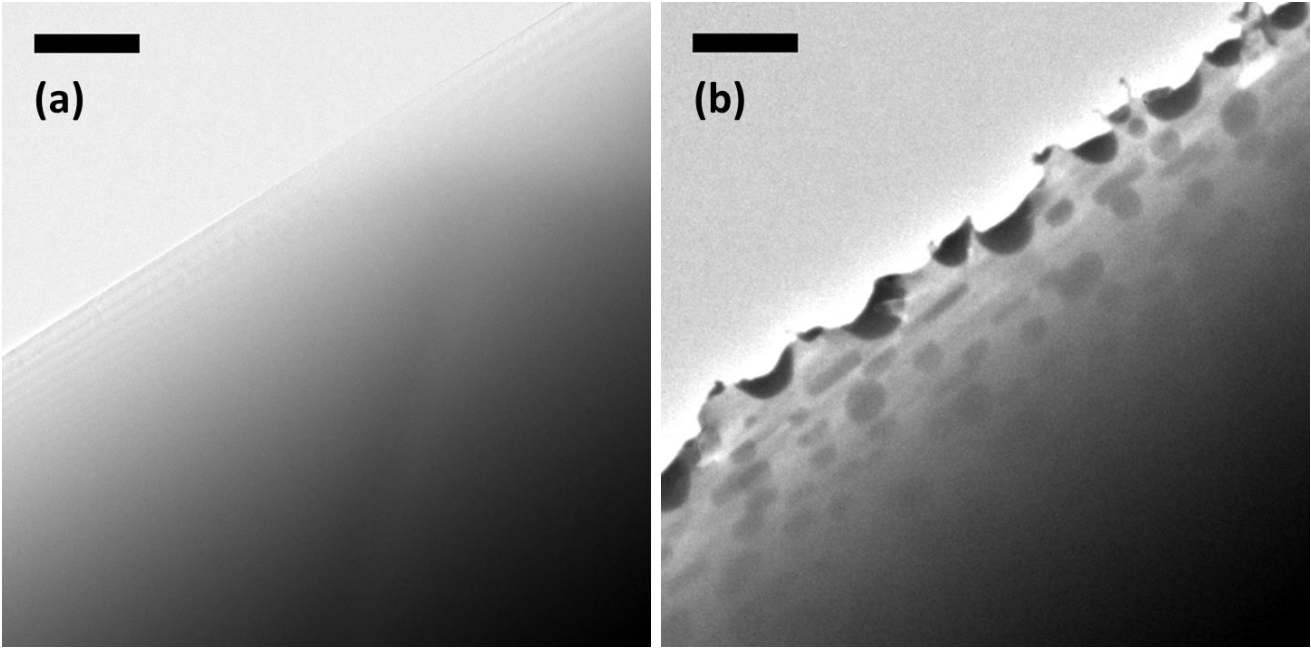

**Figure 5.** TEM images of an ion-milled GaAs thin film after brief (i.e., < 10 sec) exposure to high intensity 405 nm laser light illumination at a power density of **(a)** 19 W/cm$^2$ showing the initial degradation of the structure as evident by the bubbling visible along the diagonal thickness fringes, and at a power density of **(b)** 33 W/cm$^2$ revealing severe chemical and structural transformations as evident by the surface pitting and evolution of numerous dark and round patches of contrast. The same region of the GaAs thin film specimen is shown in both images. Scale bars correspond to 1 μm.

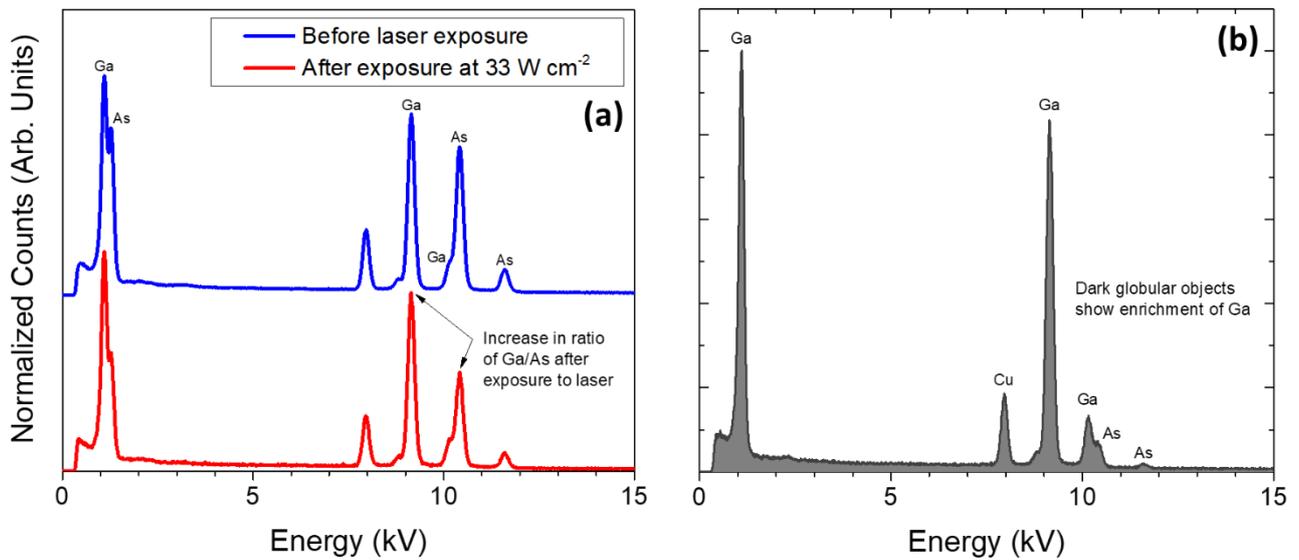

**Figure 6. (a)** EDX spectra taken of the GaAs thin film specimen before (top, blue line) and after (bottom, red line) brief exposure to 405 nm laser with power density of 33 W/cm$^2$. A comparison of the two spectra shows that there is an increase in the Ga/As ratio after the laser illumination. **(b)** An EDX spectrum acquired from one of the dark patches of contrast visible after high intensity illumination shows that they contain an enrichment of Ga, suggesting that the As in the GaAs has dissociated and sublimed while the Ga has melted into globular clusters.

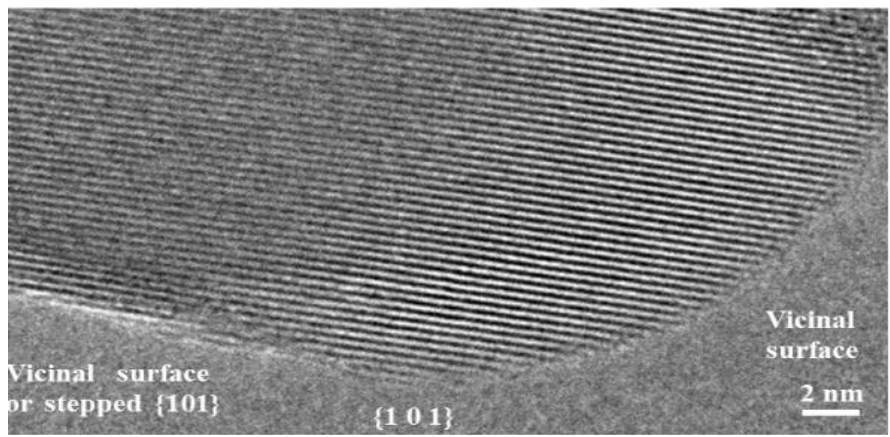

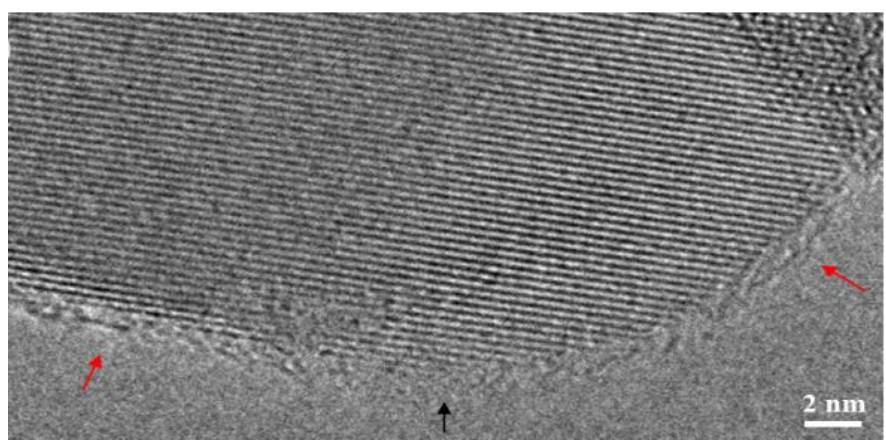

**Figure 7**. a) Image of anatase nanoparticle showing stepped (101) surface and highly stepped vicinal surface.   b) Same nanoparticle after 14 hours of exposure to light and water.

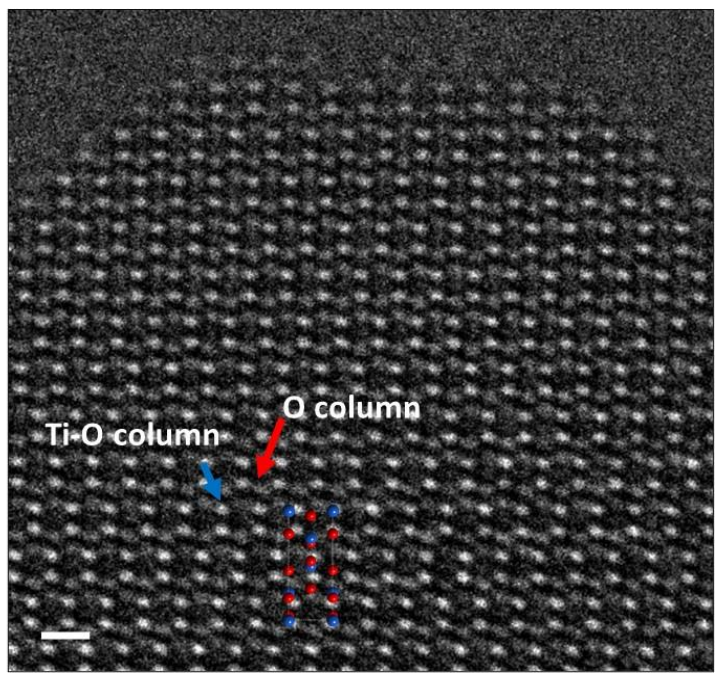

**Figure 8**. Negative Cs image of anatase nanoparticle before irradiation (scale bar 0.5 nm).

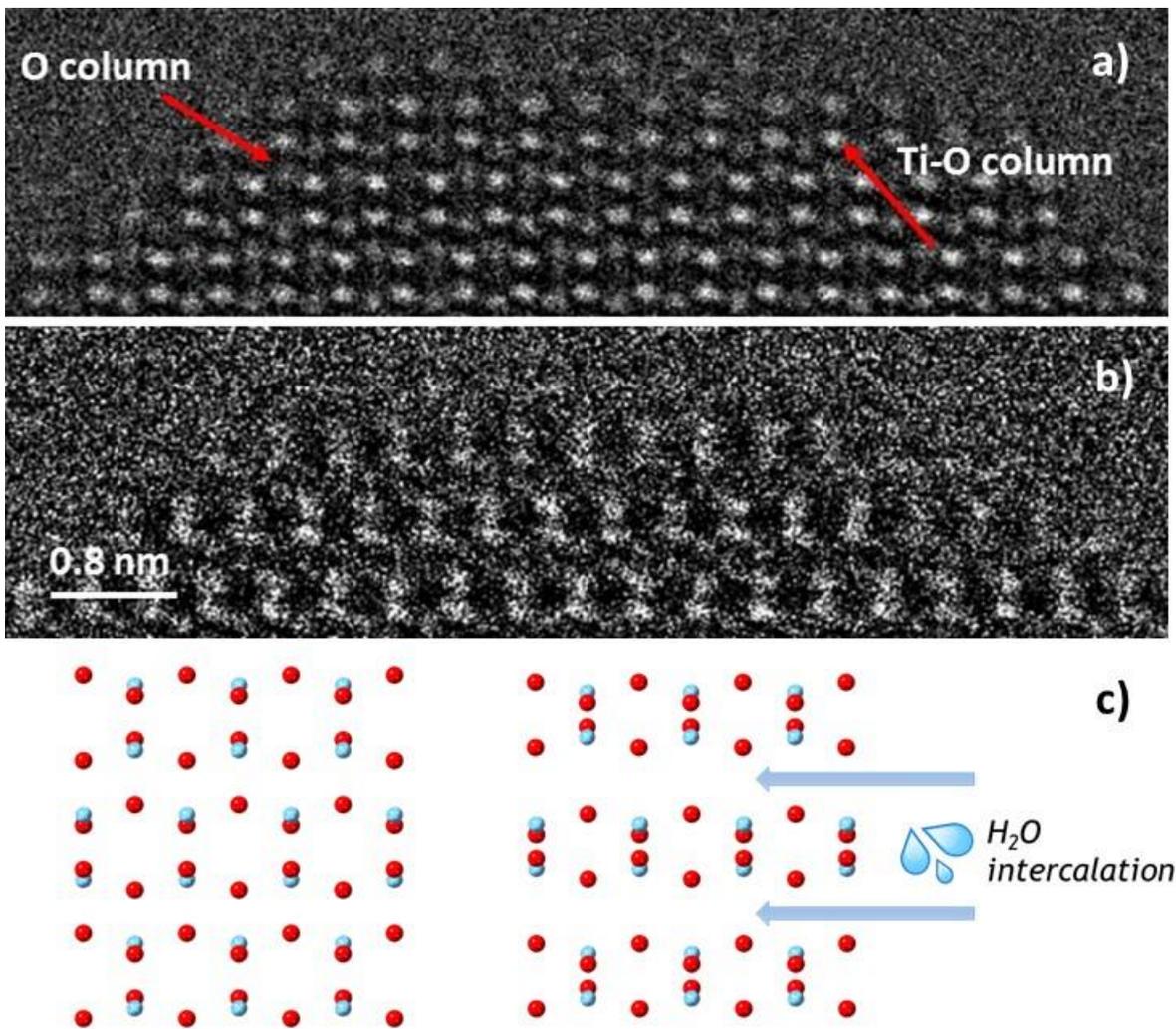

**Figure 9**. a) Image of anatase crystal before irradiation and b) after exposure to light and water. c) Schematic diagram illustrating potential intercalation mechanism.

# Supplementary Information

for

# An *In Situ* Light Illumination System for an Aberration-Corrected Environmental Transmission Electron Microscope

Qianlang Liu[1,2], Barnaby D. A. Levin[1,3], Diane M. Haiber[1,2], Joshua L. Vincent[1], and Peter A. Crozier[1]*

[1]*School for Engineering of Matter, Transport, and Energy, Arizona State University, Tempe, Arizona 85281*

[2]*Present address: Intel, Santa Clara, California, 95054*

[3]*Present address: Direct Electron, San Diego, California 92128*

*Corresponding author email: crozier@asu.edu

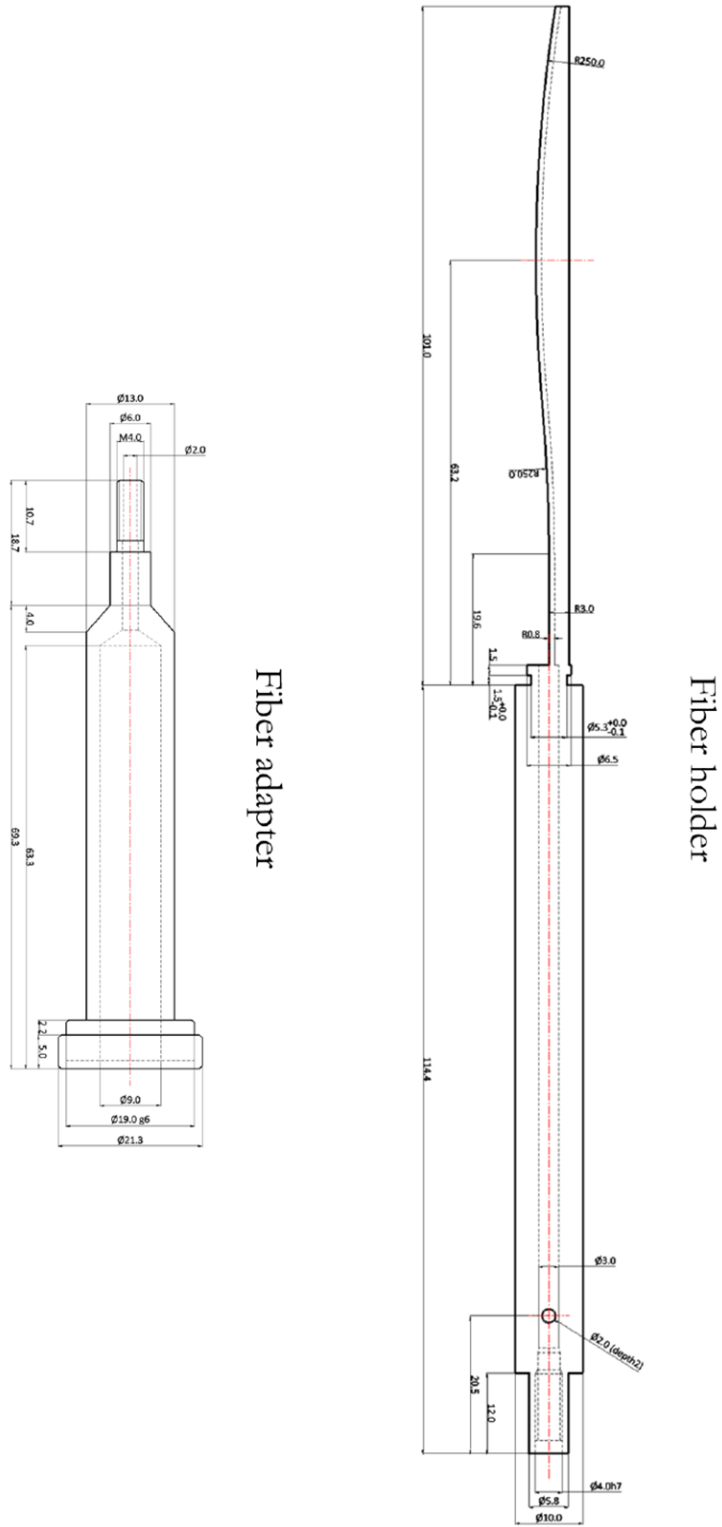

**Figure S1**. Detailed design drawings of (a) the fiber adapter and (b) the fiber holder. Measurements indicated on the drawings are in units of mm. Note the curved portion of the fiber holder, which allows the fiber to be bent when the device is inserted into the microscope.

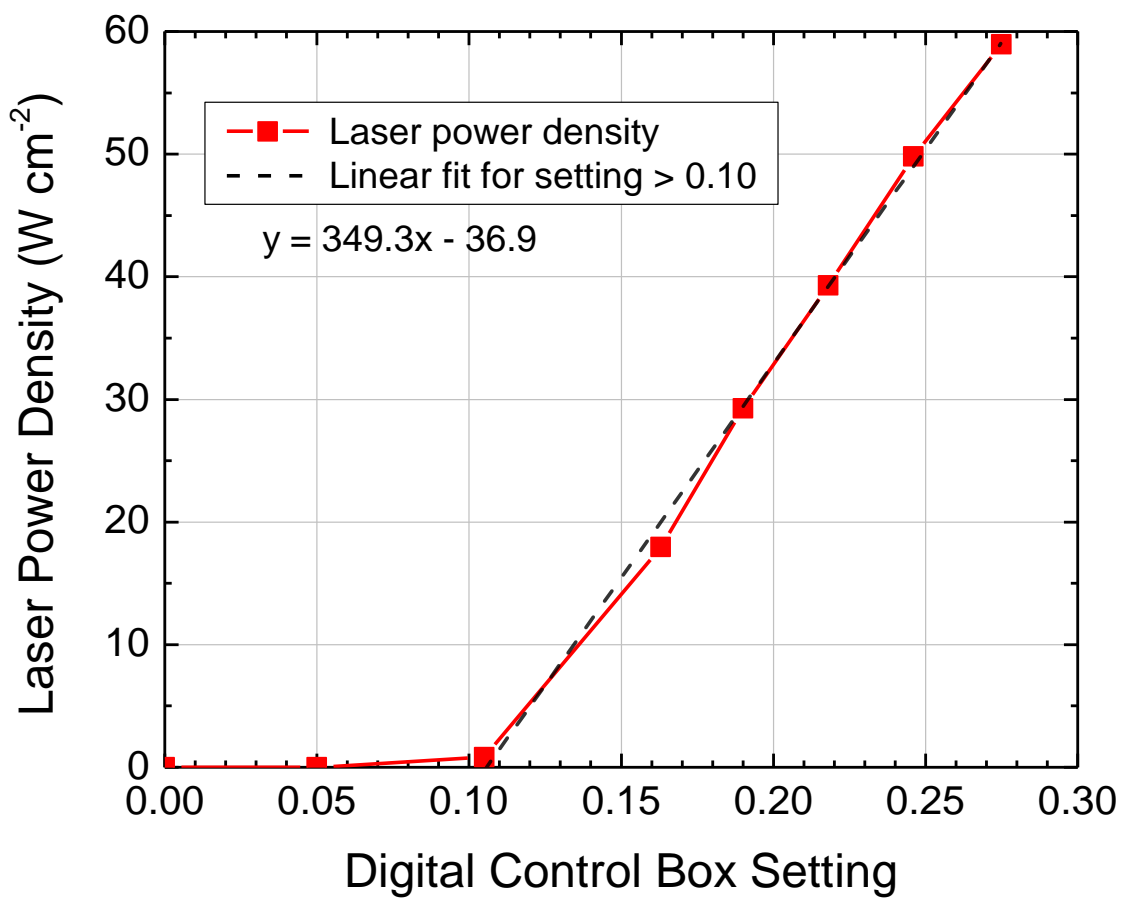

**Figure S2**. Laser power density at 3 mm plotted against the setting on the digital control box. A linear fit between the two variables is also given for control box settings > 0.10.

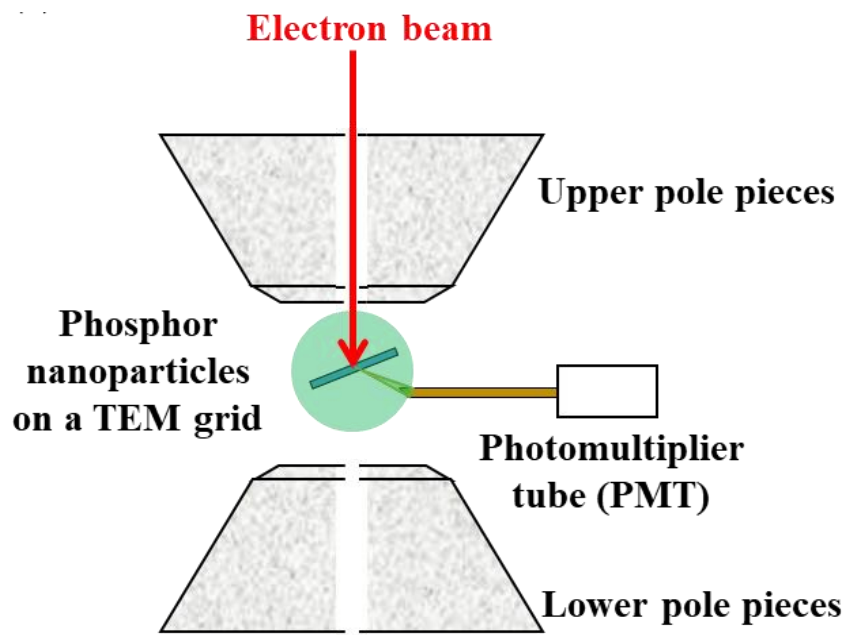

**Figure S3**: A schematic showing the principle of the alignment procedure using a photosensor system placed at the end of the fiber instead of a light source.

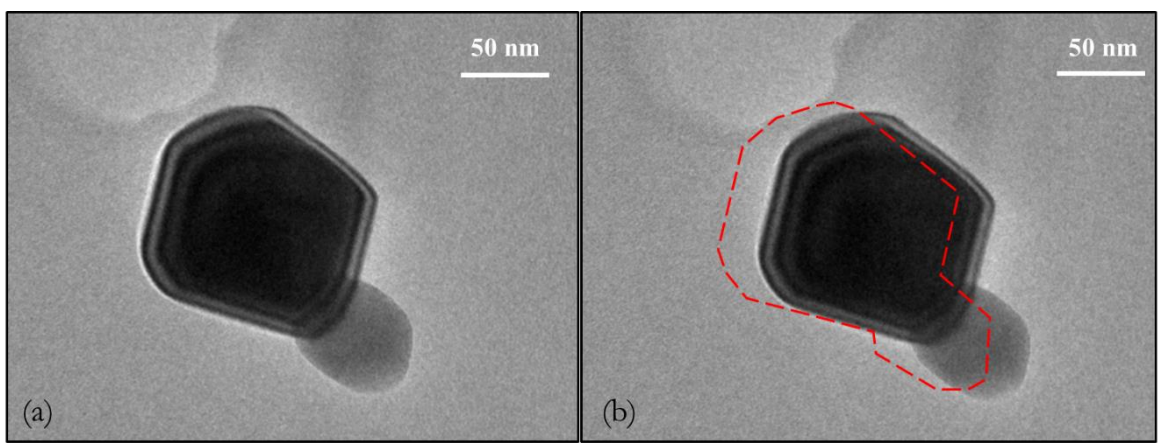

**Figure S4.** Images of the same commercial anatase cluster acquired when **(a)** minimum broadband light intensity and **(b)** maximum broadband light intensity is striking the sample area. The red dashed shape in **(b)** outlines the exact cluster location in **(a)**, which shows the image shift. The image shift is not due to mechanical or thermal instability and only occurs upon switching of the incident photo intensity.